\documentclass[%
 aps,
 amsmath,amssymb,
 reprint,%
superscriptaddress,
]{revtex4-1}

\usepackage{graphicx}
\usepackage{dcolumn}
\usepackage{bm}

\usepackage[utf8]{inputenc}
\usepackage[T1]{fontenc}
\usepackage{mathptmx}
\usepackage{etoolbox}

\usepackage{hyperref}

\makeatletter
\def\@email#1#2{%
 \endgroup
 \patchcmd{\titleblock@produce}
  {\frontmatter@RRAPformat}
  {\frontmatter@RRAPformat{\produce@RRAP{*#1\href{mailto:#2}{#2}}}\frontmatter@RRAPformat}
  {}{}
}%
\makeatother
\begin{document}


\title[A meta-GGA perspective on RuO2]{A meta-GGA perpective on the altermagnetism of RuO$_2$}

\author{Markus Meinert}
 \email{markus.meinert@tu-darmstadt.de}
\affiliation{ 
New Materials Electronics Group, Department of Electrical Engineering and Information Technology, Technical University of Darmstadt, Merckstr. 25, 64283 Darmstadt, Germany
}

\date{\today}

\begin{abstract}
The metallic oxide RuO$_2$ hosts a fascinating edge case of magnetism: while nonmagnetic in ideal bulk material, density functional theory (DFT) predicts an altermagnetic ground state within the DFT$+U$ method. The magnetic state of strained or doped thin films remains controversial, but evidence for a nontrivial magnetic state is ample. Here, I study the altermagnetic ground state of RuO$_2$ on a higher rung of Jacob's ladder of density functional approximations, the meta-GGA level including the kinetic energy density and the density Laplacian. While the workhorse functional of solid-state physics is a generalized gradient approximation (GGA), the modern r$^2$SCAN-L functional has been established as a general-purpose functional which can replace GGA, while systematically improving solid-state properties without introducing spurious errors like erroneous magnetic ground states. Comparison of LSDA+U, GGA+U, and meta-GGA+U results on RuO$_2$ shows systematic enhancement of the exchange interaction, leading to a reduction of the onset value of the Hubbard $U$ parameter at different levels of density functional approximation. However, the magnetic ground state, studied at the experimental lattice constants, remains nonmagnetic with r$^2$SCAN-L. I demonstrate that altermagnetism is easily formed upon lattice expansion, hole doping, and uniaxial strain on the c-axis. The r$^2$SCAN-L calculations set conservative thresholds for distortions and doping levels for the onset of altermagnetism in a parameter-free framework.
\end{abstract}

\maketitle

\section{Introduction}
Ruthenium dioxide (RuO$_2$) is a technologically relevant, rutile-structured metallic oxide long regarded as a Pauli–paramagnetic metal. Semilocal DFT calculations and early spectroscopy supported a nonmagnetic metallic ground state with Dirac nodal lines and a sizable intrinsic spin Hall conductivity \cite{Sun17, Jovic18, Occhialini21}. ARPES confirmed Dirac nodal lines and flat-band surface states \cite{Jovic18}, reinforcing this nonmagnetic topological-metal picture.

This view shifted when neutron diffraction indicated itinerant antiferromagnetism with a N\'{e}el temperature above room temperature \cite{Berlijn17}, and resonant x-ray scattering subsequently identified a collinear antiferromagnetic structure in both bulk crystals and thin films \cite{Zhu19}. Theory reproduced this itinerant antiferromagnetism within DFT$+U$ and many-body approaches, interpreting the antiferromagnetism as driven by a Pomeranchuk instability of the Fermi surface\cite{Berlijn17, Ahn19}. These findings positioned RuO$_2$ as a prototype for altermagnetism: a symmetry-driven momentum-dependent spin splitting in collinear antiferromagnets with zero net magnetization \cite{Smejkal20, Gonzalez21}. Predicted altermagnetic responses have been supported by thin-film observations of anomalous Hall signals \cite{Feng2022}, spin-torque generation \cite{Bose2022, Bai2022, Karube2022}, anisotropic thermal responses \cite{Zhou24}, and spectroscopic signatures of time-reversal symmetry breaking \cite{Fedchenko2024}, although other ARPES studies emphasize a paramagnetic description \cite{Liu2024}.

The resulting controversy has been intensified by muon spin rotation and renewed neutron work reporting an absence of static order in high-purity crystals \cite{Hiraishi2024, Kessler2024}. First-principles studies demonstrate that the energy difference between nonmagnetic and antiferromagnetic states is small and highly sensitive to structural parameters, doping, and correlation treatments \cite{Smolyanyuk2024}. Thin films introduce additional complexity. Epitaxial strain, interfacial bonding, and variant selection can stabilize magnetic states that are not favored in the bulk. Thin-film RuO$_2$ has been central to demonstrations of altermagnetic transport signatures \cite{Zhu19, Feng2022, Bose2022, Bai2022, Karube2022}. Recent x-ray work reports single-variant RuO$_2$(101) films with controlled N\'{e}el-vector orientation and spin-splitting magnetoresistance \cite{He2025}, underscoring the role of substrate-induced symmetry breaking.

Given these discrepancies, we attempt to clarify the robustness of altermagnetic order across different exchange–correlation functionals. Here we employ the r$^2$SCAN-L \cite{Rodriguez2020} semilocal meta-GGA and the r$^2$SCAN nonlocal meta-GGA \cite{Furness2020}, together with the PBE generalized gradient approximation (GGA) \cite{PBE} and the local spin-density approximation (LSDA) \cite{PW92}, to reassess bulk RuO$_2$ and compare systematically with DFT$+U$. Meta-GGA exchange-correlation functionals sit on the third rung of Perdew’s “Jacob’s ladder,” where the functional depends not only on the local density and its gradient but also on the kinetic-energy density or the density Laplacian, thereby allowing a more refined treatment of different bonding regimes at still semilocal cost \cite{TPSS2003}. The SCAN functional is a nonempirical meta-GGA designed to satisfy all known exact constraints appropriate to this rung and to be “appropriately normed” to key reference systems; it has been shown to describe a wide range of bonded systems and solid-state thermochemistry significantly better than standard GGAs \cite{Sun2015, Sun2016, Isaacs2018}. At the same time, several studies have demonstrated that SCAN systematically overestimates magnetic moments and magnetic energy scales in itinerant 3d transition metals, leading to “overmagnetization” in Fe, Co, Ni and related systems \cite{Isaacs2018, Ekholm2018, Fu2018}. The r$^2$SCAN functional is a regularized reformulation of SCAN that smooths its exact-constraint structure to improve numerical robustness and grid convergence while retaining most of SCAN’s accuracy \cite{Furness2020}; subsequent benchmarks for large solid-state test sets have confirmed its good performance, though the overmagnetization tendency is inherited from SCAN \cite{Kothakonda2023}. The broad success of r$^2$SCAN has led the authors and maintainers of the \textsc{Materials Project} to adopt it as the standard functional for high-throughput calculations on solids \cite{Kingsbury2022}. Deorbitalized, Laplacian-level (LL) variants such as SCAN-L and r$^2$SCAN-L remove the explicit orbital dependence by replacing the kinetic-energy density with a semilocal approximation while preserving the meta-GGA form \cite{Rodriguez2018, Rodriguez2020}. In tests on molecules and solids, r$^2$SCAN-L delivers meta-GGA-quality energetics at nearly GGA cost and largely cures the overmagnetization problem: the inflated magnetic moments of bcc Fe, hcp Co, and fcc Ni seen with SCAN and r$^2$SCAN are substantially reduced to values close to experiment in SCAN-L and r$^2$SCAN-L \cite{Rodriguez2020, Rodriguez2019}.

\section{Methods}
The calculations were performed with the \textsc{elk} code \cite{elk}, which implementes the full-potential linearized augmented plane-wave method. The meta-GGA functionals are available through an interface to the \textsc{libXC} library \cite{libxc}. The Hubbard $+U$ method was applied within the formalism of Liechtenstein et al. \cite{Liechtenstein1995}, the double-counting correction was done in the fully-localized limit, and a fixed $J=0.5$\,eV was applied. Brillouin zone sampling was performed with $9 \times 9 \times 13$ meshes. The basis set parameters were determined by the \textsc{highq} option in \textsc{elk}. The lattice constants and geometry were optimized with r$^2$SCAN-L and were found as $a = b = 4.498$\,\AA{}, $c = 3.107$\,\AA{}, and the internal position parameter of the oxygen atoms $x = 0.3048$. These values match experimental data to within 0.01\,\AA{} (0.3\%) \cite{Kiefer2025}. These parameters were used in all calculations if not stated otherwise and no additional structural relaxation was performed hereafter to achieve a clear picture of the influence of the exchange and correlation treatment in the various functionals on the stability of the altermagnetic state. For comparison, calculations with the Perdew-Burke-Ernzerhof functional \cite{PBE} on the GGA level and the Perdew-Wang functional \cite{PW92} on the local spin density approximation level were done. The muffin-tin radii of RuO$_2$ were set to 1.13\,\AA{} for Ru and 0.782\,\AA{} for O; the local magnetic moments are defined within these radii. Small, symmetry-breaking antiparallel local magnetic fields were applied within the Ru muffin-tin spheres to enable the altermagnetic ground state. Calculations on elemental solids were performed at the respective experimental lattice constants with dense Brillouin zone sampling. The r$^2$SCAN calculations were done with GPAW \cite{GPAW}, which implements the generalized Kohn-Sham framework required for self-consistent meta-GGA calculations with a nonmultiplicative potential; this is currently not available in \textsc{elk}, which only fully supports LL-meta-GGAs. The plane-wave cutoff was set to 800\,eV for convergence of magnetic moments to ca. 0.02\,$\mu_\mathrm{B}$ and the PBE PAW datasets were used with s and p electrons of Ru in valence. GPAW uses the simplified DFT$+U$ scheme of Dudarev et al.\cite{Dudarev1998}. As GPAW does not support LL-meta-GGAs, no side-by-side comparison within the same framework was done.

\begin{table}[b!]
\caption{\label{table:comparison} Comparison of magnetic moments per atom given in Bohr magneton ($\mu_\mathrm{B}$)for transition metal test cases with three density functional approximation levels. Experimental results refer to total magnetic moments including the orbital contribution, whereas the calculated values refer to spin magnetic moments only. Data from \cite{Danan1968, Myers1951}. }
\begin{tabular}{l | l l l l l}
         & Fe   & Co   & Ni   & V   & Pd   \\\hline
LSDA     & 2.18 & 1.57 & 0.59 & 0.0 & 0.0  \\
PBE      & 2.22 & 1.60 & 0.63 & 0.0 & 0.0  \\
r$^2$SCAN-L & 2.28 & 1.67 & 0.69 & 0.0 & 0.22\\
r$^2$SCAN & 2.64 & 1.74 & 0.75 & 0.0 & 0.39\\
expt. & 2.22 & 1.72 & 0.62 & 0.0 & 0.0
\end{tabular}
\end{table}

\section{results}

\subsection{Reference systems}
To begin, we first establish the accuracy of the r$^2$SCAN-L magnetic moment calculations by comparing with LSDA, PBE, and r$^2$SCAN calculations on five test cases: Fe, Co, Ni, V, and Pd. V and Pd can be seen as critical test cases, e.g., the SCAN functional finds a magnetic ground state of V. Pd is an even more sensitive test case: being a Stoner-enhanced Pauli paramagnet, it reacts easily to overestimated exchange energy densities. In Table \ref{table:comparison} we observe the general, well-known trend of $m(\mathrm{LSDA}) < m(\mathrm{PBE}) < m(\text{r$^2$SCAN-L}) < m(\text{r$^2$SCAN})$. While all three functionals correctly give V as nonmagnetic (we confirm that SCAN finds a nonzero magnetic moment), r$^2$SCAN and r$^2$SCAN-L fail the test on Pd and predict a weak magnetic moment. The magnetic moments of Fe, Co, and Ni are always a bit too high compared to accepted experimental values \cite{Reck1969}. As an additional test, we compare CoO in its NaCl-AFII structure and find that r$^2$SCAN-L ($m_\mathrm{Co} = 2.52\,\mu_\mathrm{B}$) correctly opens a band gap (as was previously shown for r$^2$SCAN \cite{Swathilakshmi2023}), whereas LSDA ($m_\mathrm{Co} = 2.38\,\mu_\mathrm{B}$) and PBE ($m_\mathrm{Co} = 2.45\,\mu_\mathrm{B}$) find a metallic ground state. This allows us to triangulate r$^2$SCAN-L as a functional which improves general structural and electronic descriptions and thermochemistry data with respect to lower-rung density functional approximations, however at the price of slightly too strong tendency towards magnetism. However, the orbital-dependent meta-GGA functional r$^2$SCAN has a substantially stronger tendency to overmagnetize solids. It was speculated that the explicit orbital dependence of the kinetic energy in r$^2$SCAN gives the functional a stronger nonlocal character, compared to the semilocal nature of the LL r$^2$SCAN-L \cite{Kaplan2022}. In metals, where the long-range Coulomb interaction is fully screened and the exchange-correlation hole is very small, the more local character of a Laplacian-level meta-GGA more accurately represents the correct exchange-correlation energy density. Therefore, r$^2$SCAN-L can be seen as a compromise between more specialized functionals for molecules, thermochemistry, and insulating solids (r$^2$SCAN) and specialized functionals for simple metallic solids (PBEsol \cite{Perdew2008}, OFR2 \cite{Kaplan2022}). As RuO$_2$ is a metallic, strongly covalent charge-transfer oxide \cite{Occhialini21}, r$^2$SCAN-L seems to be the most suitable density functional approximation. With this triangulation of functionals in mind, we use r$^2$SCAN-L as a parameter-free method to set a lower bound for the onset of altermagnetism in RuO$_2$.

\begin{figure}[t!]
\includegraphics[width=8.6cm]{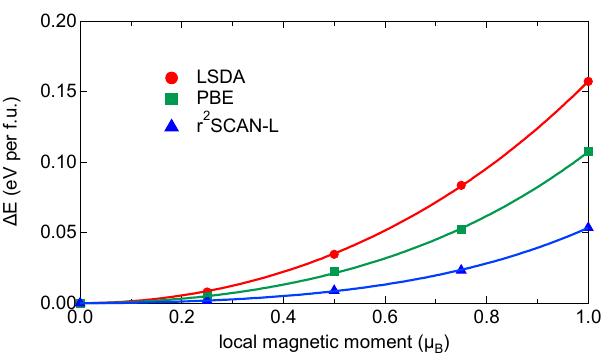}
\caption{\label{fig:E_vs_m_fsm}
Total energy change in fixed spin moment calculations. The energy difference with respect to the nonmagnetic state is given in eV per formula unit. The lines are fourth-order polynomial fits with the constant and odd terms set to zero. The fit results are $a_2(\mathrm{LSDA}) = 0.124\,\mathrm{eV}/\mu_\mathrm{B}^2$, $a_2(\mathrm{PBE}) = 0.076\,\mathrm{eV}/\mu_\mathrm{B}^2$, $a_2(\mathrm{r^2}\text{SCAN-L}) = 0.018\,\mathrm{eV}/\mu_\mathrm{B}^2$.}
\end{figure}

\begin{figure}[t!]
\includegraphics[width=8.6cm]{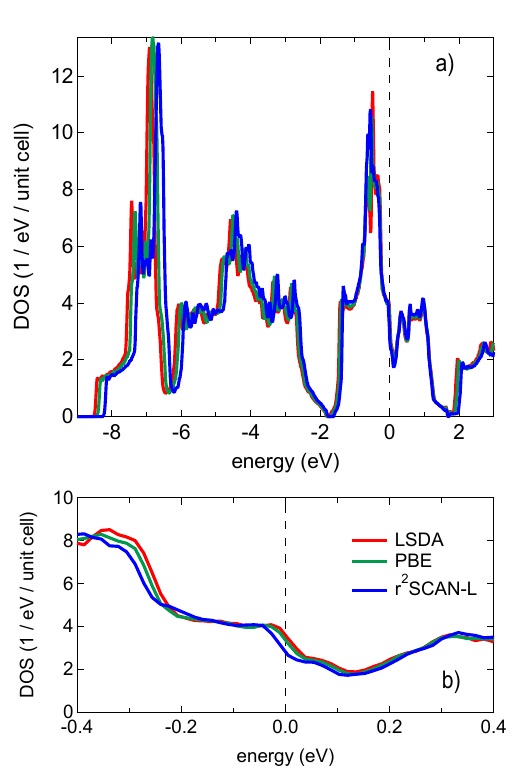}
\caption{\label{fig:DOS}
Density of states plot for LSDA, PBE, and r$^2$SCAN-L functionals. a) is the full plot, b) is a close-up view around the Fermi energy.}
\end{figure}

\subsection{Groundstate of RuO$_2$}
The semilocal functionals LSDA, PBE, and r$^2$SCAN-L find a nonmagnetic groundstate for RuO$_2$. In Figure \ref{fig:E_vs_m_fsm} we show the energy difference with respect to this groundstate in fixed spin moment calculations, where the Ru magnetic moments are set antiparallel inside the two muffin-tin spheres of the unit cell. A clear hierarchy of the three functionals is seen, where LSDA has the largest energy increase and r$^2$SCAN-L has the smallest energy increase with increasing local magnetic moments. We expand the energy in a Landau series, $\Delta E(m_\mathrm{Ru}) = a_2 m_\mathrm{Ru}^2 + a_4 m_\mathrm{Ru}^4 + ...$  and fit the energy curves to fourth order. The quadratic term $a_2$ directly gives a measure for the proximity to the magnetic instability, with the instability criterion being $a_2 = 0$. We now apply an effective local Stoner model as an effective mapping of the local exchange interaction due to the different types of functionals on the Ru site: we have for the interaction term $E_\mathrm{ex} = -\frac{1}{4} I_\mathrm{Ru} m_\mathrm{Ru}^2$ with the instability criterion $I_\mathrm{Ru}N_\mathrm{Ru}(E_\mathrm{F}) = 1$ and therefor $a_2 = \frac{1}{4N_\mathrm{Ru}(E_\mathrm{F})} - \frac{I_\mathrm{Ru}}{4}$, where $I_\mathrm{Ru}$ is the effective Ru-site Stoner exchange parameter and $N_\mathrm{Ru}(E_\mathrm{F})$ is the Ru site-projected density of states per spin (DOS). Within this effective mapping, trends in $I_\mathrm{Ru}$ directly reflect changes in both Ru $4d$ localization and Ru-O hybridization across the functional hierarchy. With the calculated densities of states, we can obtain estimates for the effective Ru-site Stoner exchange parameters. In this way, the effective local Stoner criterion is directly connected to the DFT fixed-moment energy curve. With $N_\mathrm{Ru}(E_\mathrm{F})$ being 0.561\,eV$^{-1}$ (LSDA), 0.524\,eV$^{-1}$ (PBE), 0.443\,eV$^{-1}$ (r$^2$SCAN-L), we obtain for the local Stoner parameter: 0.80\,eV (LSDA), 1.3\,eV (PBE), and 2.12\,eV (r$^2$SCAN-L). We point out that the systematic reduction of bandwidths pushes the Ru $4d$ states around the Fermi energy to lower energies, thus reducing the DOS in this region (Fig. \ref{fig:DOS} b)). Thus, the hierarchy of functionals increases systematically the local exchange interaction on the Ru atoms. The r$^2$SCAN functional finds the altermagnetic state as the groundstate with magnetic moments of $\pm 0.98\,\mu_\mathrm{B}$. It correspondingly assigns an even larger effective Stoner parameter to the Ru site. As the higher-rung functionals reduce the self-interaction error of LSDA (which results in artificial delocalization of electrons), the electrons are more localized and less hybridized as we progress from LSDA via GGA to meta-GGA functionals. This is confirmed by the narrowing bandwidths in the density of states plots (Fig. \ref{fig:DOS} a)), and by inspecting the number of interstitial electrons (electrons in the regions between the muffin-tin spheres), which reduces from 10.31 in LSDA, to 10.18 in PBE, and further to 9.97 in r$^2$SCAN-L. These electrons localize on the oxygen atoms, whereas the charges in the Ru muffin-tin spheres remain nearly unchanged.  Based on the previous assessment of the tendency of r$^2$SCAN-L towards overmagnetization in itinerant systems, while this functional still finds a nonmagnetic groundstate in RuO$_2$, we thus conclude that the groundstate of RuO$_2$ is correctly identified as nonmagnetic. This is in agreement with a broad consensus of experiments on stoichiometric bulk material \cite{Kiefer2025}.

\begin{figure}[t!]
\includegraphics[width=8.6cm]{m_vs_U}
\caption{\label{fig:m_vs_U}
Local magnetic moments on the Ru site as a function of the Hubbard $+U$ parameter for different exchange-correlation functionals.}
\end{figure}

In Figure \ref{fig:m_vs_U}, the DFT$+U$ results for the Ru local magnetic moments are shown, comparing LSDA$+U$, PBE$+U$, r$^2$SCAN-L$+U$, and r$^2$SCAN$+U$. A clear progression of the magnetic moments as a function of the $+U$ parameter with higher rung functionals is seen. For LSDA, the onset of altermagnetism is at $U \approx 2.2$\,eV, for PBE the onset is at $U \approx 1.2$\,eV, and for r$^2$SCAN-L the onset is found at $U \approx 0.2$\,eV. r$^2$SCAN finds the altermagnetic groundstate without addition of the Hubbard $U$, and further enhances the moments upon adding the Hubbard term. The curves are identical up to a shift on the horizontal axis. Our result for PBE$+U$ is very similar to a previous PBE$+U$ study reporting the onset at $U=1.06$\,eV \cite{Smolyanyuk2024}. DFT$+U$ increases magnetic moments because the Hubbard term penalizes partial occupancies and, in particular, intra-orbital double occupancy of the correlated $d$ orbitals, driving them toward integer occupancies. This enhances the exchange splitting and reduces artificial delocalization, thus favoring parallel spins in different orbitals and enhancing local moments. 

By combining fixed-moment Landau fits with DFT$+U$ calculations, we find that the quadratic coefficient of the magnetic energy, $a_2(U)$, obeys a linear relation $a_2(U) = a_2 - cU$ with a functional-independent slope $c \approx 0.06\,\mu_\mathrm{B}^{-2}$. By inserting the negative $U$ for the magnetic instability in the r$^2$SCAN calculation, we determine the Stoner parameter as $I_\mathrm{Ru} = 2.64$\,eV for this functional. This appears, expectedly, like a gross overestimate: typical values for $4d$ transition metals, and Ru in particular, are expected around $I \approx 0.5$\,eV, whereas values around $I \approx 1$\,eV are typically seen in late $3d$ transition metals based on LSDA calculations \cite{Fritsche1998}. LSDA, PBE, and r$^2$SCAN-L differ only in the ‘bare’ curvature $a_2$, that is, in their intrinsic distance from the magnetic instability, while the efficiency with which the Hubbard term drives the system toward altermagnetism is the same. This indicates that the DFT$+U$ correction effectively renormalizes a local Ru-centered Stoner parameter in a manner that is controlled primarily by the Ru 4d manifold and the chosen correlated subspace, rather than by the underlying exchange–correlation functional.

It was argued previously that values of $U$ beyond 1\,eV appear too large for PBE-based calculations on RuO$_2$ because of the metallic screening and the strong hybridization, which should strongly diminish $U$. Values between 1\,eV and 3\,eV were seen as more appropriate for the Mott insulators  RuCl$_3$, RuBr$_3$, and RuI$_3$ \cite{Smolyanyuk2024, Kaib2022}.

\begin{figure}[t!]
\includegraphics[width=8.6cm]{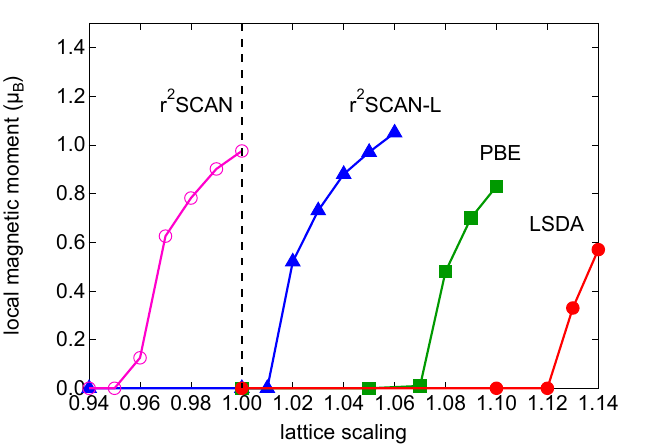}
\caption{\label{fig:m_vs_scale}
Local magnetic moments on the Ru site as a function of the isotropic lattice scaling parameter for different exchange-correlation functionals.}
\end{figure}

\subsection{Lattice distortion and doping}

The lattice volume plays a critical role for magnetic materials, where larger lattice volume localizes electrons, reduces hybridization and screening, and thus makes the orbitals more atomic-like. This enhancement of on-site exchange leads to an increase of the Stoner parameter, while also enhancing $N_\mathrm{Ru}(E_\mathrm{F})$. We see this trend in RuO$_2$, where the onset of altermagnetism follows a similar pattern as for the Hubbard $U$. With just 2\% increase of lattice scaling (an isotropic expansion of all lattice constants), altermagnetism is found for r$^2$SCAN-L, whereas 8\% is required for PBE, and 11\% is needed for LSDA. A compression of the lattice constants by about 5\% suppresses the altermagnetic groundstate with r$^2$SCAN. Evaluating $N_\mathrm{Ru}(E_\mathrm{F})$ at the onset expansions, we find the effective Ru-site Stoner parameters are enhanced to 1.44\,eV (LSDA, $N_\mathrm{Ru}(E_\mathrm{F})=0.69$eV$^{-1}$);  1.8\,eV (PBE, $N_\mathrm{Ru}(E_\mathrm{F})=0.56$eV$^{-1}$);  2.24\,eV (r$^2$SCAN-L, $N_\mathrm{Ru}(E_\mathrm{F})=0.45$eV$^{-1}$). Thus, r$^2$SCAN-L puts RuO$_2$ at the verge of being an altermagnet and minimal lattice distortion may already lead to a magnetic groundstate. 

\begin{figure}[t!]
\includegraphics[width=8.6cm]{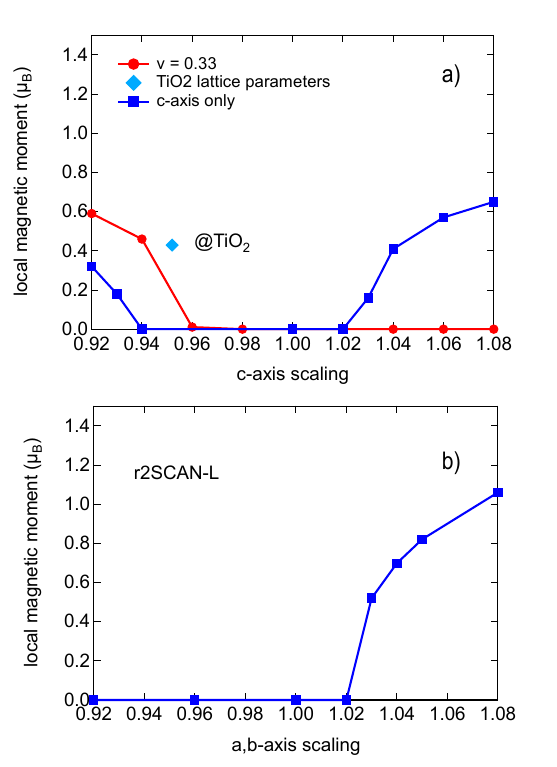}
\caption{\label{fig:m_vs_strain}
Local magnetic moments on the Ru site as a function of uniaxial strain for a) c-axis only scaling, strain including the Poisson ratio, and for the TiO$_2$ lattice parameters; b) for a,b-axis only scaling. All results were obtained with the r$^2$SCAN-L functional.}
\end{figure}

Uniaxial strain is studied with r$^2$SCAN-L and is considered in three ways: $c$-axis strain only, $a,b$-axis strain only, and $c$-axis strain with $a$ and $b$ parameters given by the Poisson number $\nu =0.33$ of RuO$_2$ \cite{osti_1307989}, where $- \nu \mathrm{d}c/c = \mathrm{d}a/a = \mathrm{d}b/b$. In Figure \ref{fig:m_vs_strain} a), we observe that $c$-axis only strain enables altermagnetism both ways, where a $c$-axis expansion of 3\% or compression of -7\% are the onsets. For in-plane strain (i.e. a- and b-axis scaling), formation of magnetic moments is observed at 3\% expansion of both the a- and b-axis. Under compression, the nonmagnetic state remains stable. Taking the Poisson ratio into account, we find that c-axis expansion does not provide altermagnetism anymore, only the c-axis compression with simultaneous $a$,$b$-axis expansion allows altermagnetism. This resembles the structure of rutile TiO$_2$ ($a=4.59$\,\AA{}, $c=2.96$\,\AA{}), a commonly used substrate for RuO$_2$ epitaxial film growth: at the experimental TiO$_2$ lattice constants, a magnetic moment of 0.4\,$\mu_\mathrm{B}$ is found. With the Poisson ratio taken into account, the lattice volume is expanded with c-axis scaling $> 1$, but the a and b axis shrink, the effect of which appears to outcompete the lattice volume increase and no magnetic moment forms. The onset of altermagnetism with the Poisson ratio included is found with a higher scaling factor compared to c-axis only scaling because of the simultaneous a- and b-axis expansion.

\begin{figure}[t!]
\includegraphics[width=8.6cm]{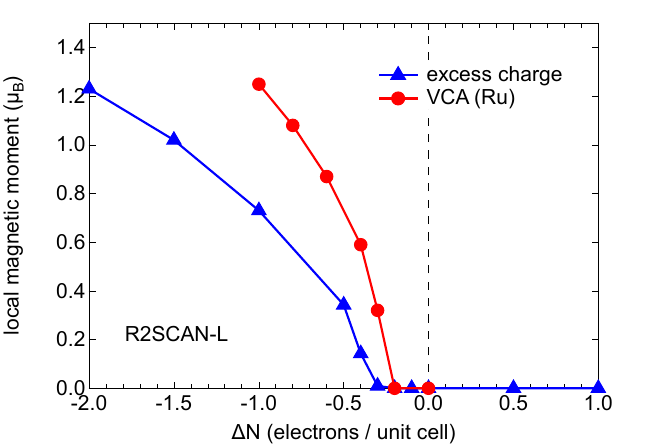}
\caption{\label{fig:m_vs_excesscharge}
Local magnetic moments on the Ru site obtained with r$^2$SCAN-L for unit cell doping via excess charge per unit cell and within the virtual crystal approximation (VCA) on the Ru site.}
\end{figure}

Finally, the doping of RuO$_2$ is investigated with r$^2$SCAN-L in two ways: a) as a global, homogeneous excess charge of the unit cell; and b) via the virtual crystal approximation (VCA) by changing the nuclear charge of Ru and correspondingly changing the number of electrons. The former case corresponds to a rigid-band Fermi energy shift, whereas the latter case may be considered as the more realistic model, which treats isolated Ru vacancies, more realisitically. In both models, formation of magnetic moments is observed, which is readily explained by the increase of $N_\mathrm{Ru}(E_\mathrm{F})$ below the Fermi energy. In the VCA, the onset of altermagnetism is observed at $-0.1$ electrons per Ru atom, whereas the homogeneous excess charge model indicates closer to $-0.2$ electrons per Ru atom ($-0.4$ electrons per unit cell) as the onset. This result indicates that a modest doping with early $3d$ or $4d$ transition metals, e.g. Ti, may induce altermagnetism: each Ti atom reduces the valence charge by four electrons, thus $1/40 = 2.5\%$ of Ti substituting for Ru is sufficient to achieve a hole doping of $-0.1$ electrons per cation site. A substantial magnetic moment around $1\,\mu_\mathrm{B}$ per Ru atom would be achievable with a 10\% Ti substitution. 

\section{Conclusion}
In this study, we investigated the altermagnetic instability of RuO$_2$ using the r$^2$SCAN-L meta-GGA functional. We interpret our results in within a local effective Stoner model and demonstrate that higher-rung functionals effectively increase the on-site Stoner exchange parameter. Because r$^2$SCAN-L retains the overmagnetization tendency characteristic of meta-GGAs for itinerant metals, it provides a conservative lower bound on the lattice distortions and hole-doping levels required to stabilize altermagnetism; within this framework the bulk ground state of RuO$_2$ remains nonmagnetic. These thresholds offer a useful baseline for interpreting thin-film experiments that report altermagnetic signatures: such films must depart sufficiently from the ideal bulk structure, through strain, stoichiometry, or carrier doping, to cross the onset conditions identified here. When several mechanisms act simultaneously--hole doping from intermixing at the interface, epitaxial tensile strain in the $ab$ plane, and possibly magnetic proximity--an interface-localized altermagnetic state becomes plausible and provides a consistent explanation for the observed transport phenomena.

\begin{acknowledgments}
We acknowledge the many open-source software developers and contributors whose efforts make advanced electronic-structure research broadly accessible. Their dedication to creating, maintaining, and improving these tools has been essential for enabling the present work.

\end{acknowledgments}

\section*{AUTHOR DECLARATIONS }

\subsection*{Conflict of Interest}
The authors have no conflicts to disclose.

\section*{Data Availability Statement}

The data that support the findings of this study are available upon reasonable request from the corresponding author.

\section*{References}

\bibliography{cite} 

\end{document}